\documentclass[secnumarabic,twocolumn,aps,showpacs,showkey,superscriptaddress,prl,floatfix]{revtex4-1}

\usepackage{amsmath,amssymb,amsthm}

\usepackage{hyperref}
\usepackage[dvips]{graphicx}
\usepackage{amsfonts}
\usepackage{hyperref}
\usepackage{bbm}
\usepackage{float}
\usepackage{color}
\usepackage{mdwlist}
\usepackage{subfigure}
\usepackage[normalem]{ulem}

\definecolor{orange}{rgb}{1.00,0.50,0.0}

\newcommand{\C}{\mathbb{C}}
\renewcommand{\H}{\mathcal{H}}
\newcommand{\1}{\mathbbm{1}}

\newcommand{\E}{\mathbb{E}}

\renewcommand{\O}[1]{\mathcal{O}\left(#1\right)}
\renewcommand{\o}[1]{o\left(#1\right)}

\newcommand{\mr}{\mu_{R}}

\newcommand{\iid}{\emph{iid}~}
\newcommand{\im}{\emph{i}~}
\newcommand{\h}{\emph{ht}~}

\graphicspath{{../figures/}}

\begin{document}

\title{Real eigenvalues of non-symmetric random matrices: Transitions and Universality}

\author{Luis Carlos Garc\'ia del Molino}
\email{garciadelmolino@ijm.univ-paris-diderot.fr}
\affiliation{Institut Jacques Monod, CNRS UMR 7592, Universit\'e Paris Diderot, Paris Cit\'e Sorbonne, F-750205, Paris, France}
\affiliation{Mathematical neuroscience Team, CIRB-Coll\`ege de France\footnote{CNRS UMR 7241,
INSERM U1050, UPMC ED 158, MEMOLIFE PSL*} and INRIA Paris-Rocquencourt, MYCENAE Team, 11 place Marcelin Berthelot, 75005 Paris, France}

\author{Khashayar Pakdaman}
\affiliation{Institut Jacques Monod, CNRS UMR 7592, Universit\'e Paris Diderot, Paris Cit\'e Sorbonne, F-750205, Paris, France}

\author{Jonathan Touboul}
\affiliation{Mathematical neuroscience Team, CIRB-Coll\`ege de France\footnote{CNRS UMR 7241,
INSERM U1050, UPMC ED 158, MEMOLIFE PSL*} and INRIA Paris-Rocquencourt, MYCENAE Team, 11 place Marcelin Berthelot, 75005 Paris, France}

\date{\today}%
\maketitle

\textbf{In the past 20 years, the study of real eigenvalues of non-symmetric real random matrices has seen important progress \cite{Edelman94,Edelman97,kanzieper05,Forrester07,akemann07,kanzieper2015,garcia-JSP:16}. Notwithstanding, central questions still remain open, such as the characterization of their asymptotic statistics and the universality thereof. 
In this letter we show that for a wide class of matrices, the number $k_n$ of real eigenvalues of a matrix of size $n$ is asymptotically Gaussian with mean $\bar k_n=\O{\sqrt{n}}$ and variance $\bar k_n(2-\sqrt{2})$. Moreover, we show that the limit distribution of real eigenvalues undergoes a transition between bimodal for $k_n=\o{\sqrt{n}}$ to unimodal for $k_n=\O{n}$, with a uniform distribution at the transition. We predict theoretically these behaviours in the Ginibre ensemble using a log-gas approach, and show numerically that they hold for a wide range of random matrices with independent entries beyond the universality class of the circular law. 
}

Random matrices have become a central tool in the modelling of large-scale interacting systems. Applications range from nuclear and statistical physics~\cite{wigner1955lower,auffinger2013random} to ecology~\cite{may1972will} and neuroscience~\cite{sompolinsky-crisanti-etal:88}. An explosion of works in physics and mathematics have revealed in the past few years their astounding universal properties~\cite{Bordenave12,Bourgade14a,Tao10}. 
Real non-symmetric random matrices have the surprising property that an unbounded number of eigenvalues accumulate on the real axis. Understanding the respective distributions of real and non-real eigenvalues of the spectrum is very important for applications in different fields. For instance, in certain superconducting systems, the statistics of level crossings are identical to those of real eigenvalues of real non-symetric matrices \citep{Beenakker13}. Another example are biologically inspired neuronal networks where the real or non-real nature of leading eigenvalues governs the global dynamics~\cite{garcia:13}

A number of important results were already established in the classical case of the real Ginibre ensemble (matrices with independent Gaussian elements with variance $1/n$). The expected number $\bar{k}_n$ of real eigenvalues of $n\times n$ real Ginibre matrices diverges as $\sqrt{2n/\pi}+1/2+o(1)$~\cite{sommers88,Edelman94} and the variance of the number $k_n$ of real eigenvalues is  $\Sigma^2_n=(2-\sqrt{2})\bar{k}+o(\sqrt{n})$ \cite{Forrester07}. At leading order, these are universal properties of matrices with independent elements with exponentially decaying distribution and moments matching the normal distribution up to fourth order~\cite{Tao10}. The distribution $p^n_k$ of $k_n$ was characterized formally in~\cite{Edelman97,kanzieper05} for the real Ginibre ensemble, and estimated numerically for small $n$, or analytically for $k_n=n$, $n-2$ and $0$~\cite{akemann07,kanzieper2015}, and the typical behavior characterized for $k_n=\O{n}$~\cite{garcia-JSP:16}.

Beyond the statistics of the number of eigenvalues, the averaged distribution of real eigenvalues of the Ginibre ensemble was shown to be asymptotically uniform on the interval $[-1,1]$~\cite{Edelman94}, with local statistics close to semi-Poisson~\cite{Beenakker13,Forrester13}. However, non-averaged properties of the distribution of real eigenvalues have not been studied. 

In this Letter we explore the distribution $p^n_k$ and the distribution of real eigenvalues for large $n$ and find that they exhibit a rich and, to some extent universal, phenomenology. 

Our first finding is that for large Ginibre matrices the distribution of $k_n$ is Gaussian. This result stems from the central limit theorem and relies on the short correlation distances between real eigenvalues. This is shown using the \emph{log-gas analogy} that allows to identify eigenvalues to electrostatic particles in a confining potential $W$ (see Methods). We also obtain that the number of real eigenvalues in an interval $[-1,x]$ has Gaussian independent increments and therefore, when properly normalized, it is the Brownian motion on $[-1,1]$. These results are a major refinement on the characterization of the number of real eigenvalues in the Ginibre ensemble. Furthermore, since they are a consequence of the local properties of the spectrum, they hold for the matrices in the universality class of~\cite{Tao10}. 

\begin{figure}[t]
\centering
 \includegraphics[width=0.48\textwidth]{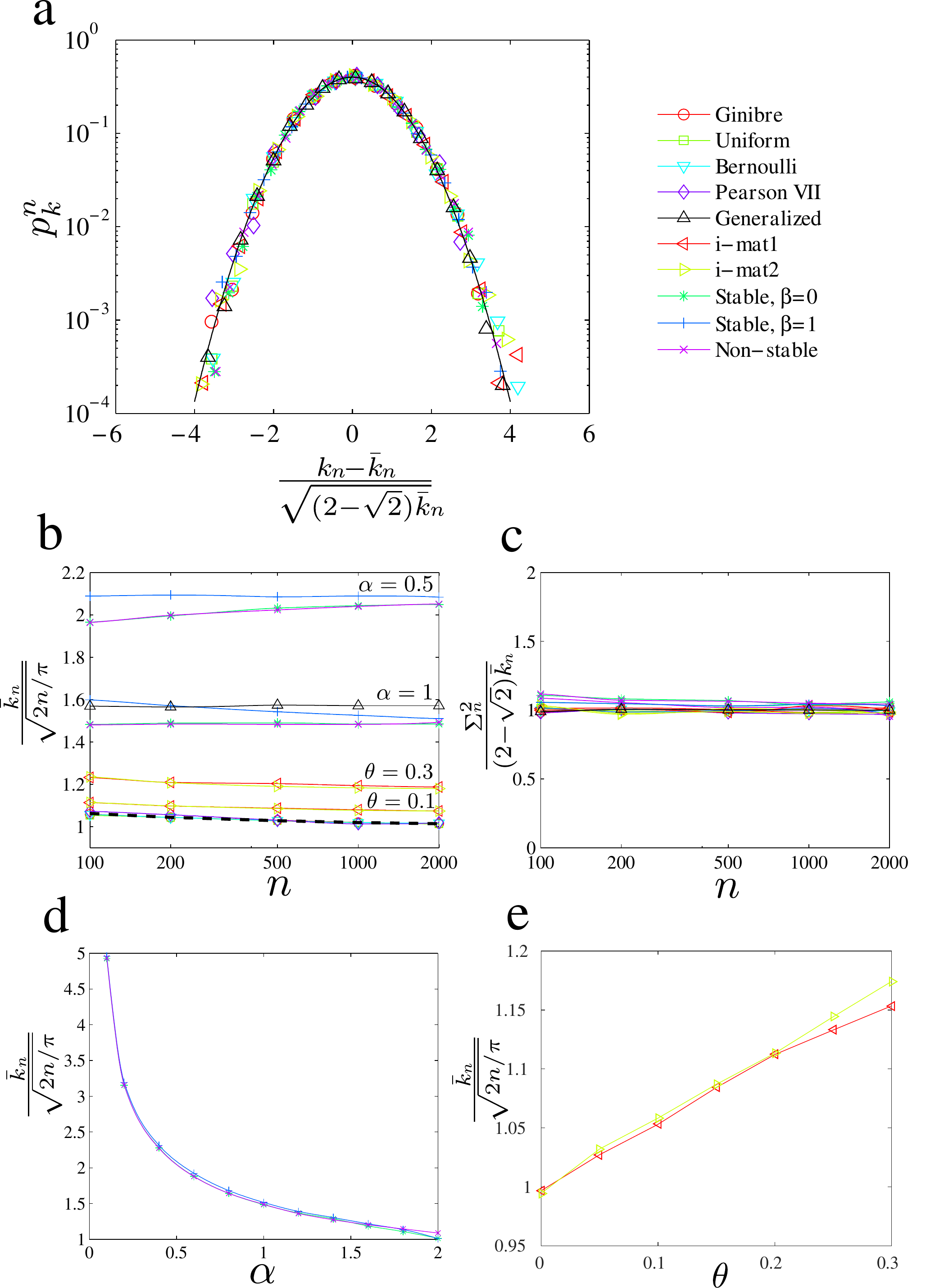}
 \caption{\small \textbf{Statistics of $k_n$. (a)} Rescaled $p^n_k$ for $n=1000$ and 9 families of matrices precisely collapse on the normal distribution (black line). The families studied include matrices with entries with unbounded second moment, matrices with non identically distributed entries and even generalized eigenvalues. \textbf{(b, c)} Numerical estimations for the mean and variance of $k_n$ for 10 different families of matrices. The dashed line corresponds to the expected value for Ginibre $1+\frac{1}{2\sqrt{2n/\pi}}$. \textbf{(d, e)} Dependence of $\bar k_n$ on $\alpha$ and $\theta$ for matrices with unbounded second moment and non identically distributed entries respectively with $n=1000$. We observe that the limit of $\bar k_n/\sqrt{n}$ is minimal for independently identically distributed (\iid) matrices and for all other cases it depends only and monotonically on $\alpha$ and $\theta$ respectively. Furthermore, the statistics converge to the values found for \iid matrices as $\alpha^-\to2$ and $\theta^+\to0$ (i.e. when the matrices converge themselves to \iid matrices). See Methods for details about the statistics and generation of the matrices.\vspace{-2mm}}\label{fig:pnk}%
\end{figure}

Through intensive numerical simulations, we confirm these results and observe that they hold for matrices far beyond the universality class of~\cite{Tao10}. In particular we find that despite possible variations between matrix classes, the distribution of $p^n_k$ is always asymptotically Gaussian (figure.~\ref{fig:pnk} (a)). This is also the case of the scaling of $\bar{k}_{n}\sim c\, \sqrt{n}$ and of the asymptotic ratio $\Sigma_n^2/\bar k_n \to (2-\sqrt{2})$. Indeed, figure \ref{fig:pnk}(b, c) suggests that these hold for matrices with (i) non identically distributed elements (ii) with unbounded second moment. We further observe that the coefficient $c$ only depends on simple statistics of the law of their elements in case (i), and only on the tail of the elements distribution in case (ii).

In order to further characterize the universality of these observations, we show in the Supplementary Material that for matrices with strongly correlated transposed entries, or products of many matrices, the scaling of $\bar{k}_{n}$ changes and may take any value. However, even for those matrices, $p^n_k$ is still Gaussian and, strikingly, the limit property $\Sigma_n^2/\bar k_n \to (2-\sqrt{2})$ persists, revealing a very robust property of the statistics of real eigenvalues. Along the same lines, we analysed numerically the properties of the integer-valued stochastic process $k_n(t)$ of the number of real eigenvalues of simple matrix valued Gaussian stochastic processes, and found that it converges to Gaussian white noise as $n\to\infty$ (see Supp. Mat.).

\begin{figure}[b]
\centering
\includegraphics[width=.5\textwidth]{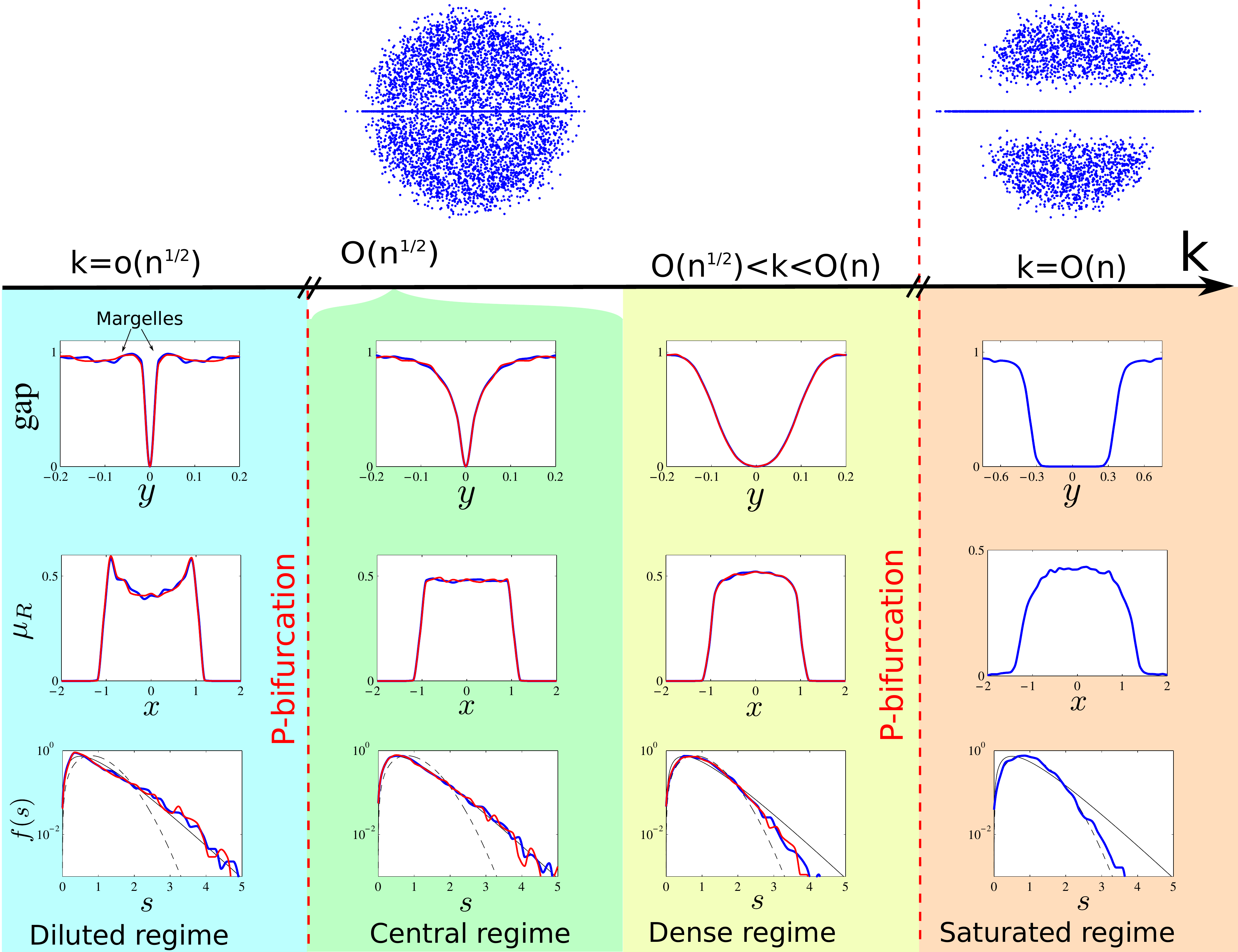}
 \caption{\small  \textbf{Transition of $\mr$ for \iid matrices.} Statistics for $5000$ Ginibre (blue), centered uniform (red) random matrices with $n=200$. In the saturated regime we use simulations of the log-gas instead of matrices. First row: The macroscopic distribution converges to the circular law except for the saturated regime. Second row: profile of the gap around the real axis, it is clear how it increases as $k_n$ increases. In the diluted regime we can see that the excess mass removed from the real axis concentrates close to it forming \emph{margelles} at both sides of the axis. Third row: real distribution $\mr$, shows a clear transition from bimodal to unimodal. Bottom row: Normalized spacing distributions compared to semi-Poisson ($4se^{-2s}$, solid line) and Wigner surmise ($\frac{\pi}{2}se^{-\pi s^2/4}$, dashed line). The coincidence between Ginibre and uniform matrices is remarkable. \vspace{-2mm}}\label{fig:main}%
\end{figure}

\begin{figure}[t]
\centering
\includegraphics[width=.48\textwidth]{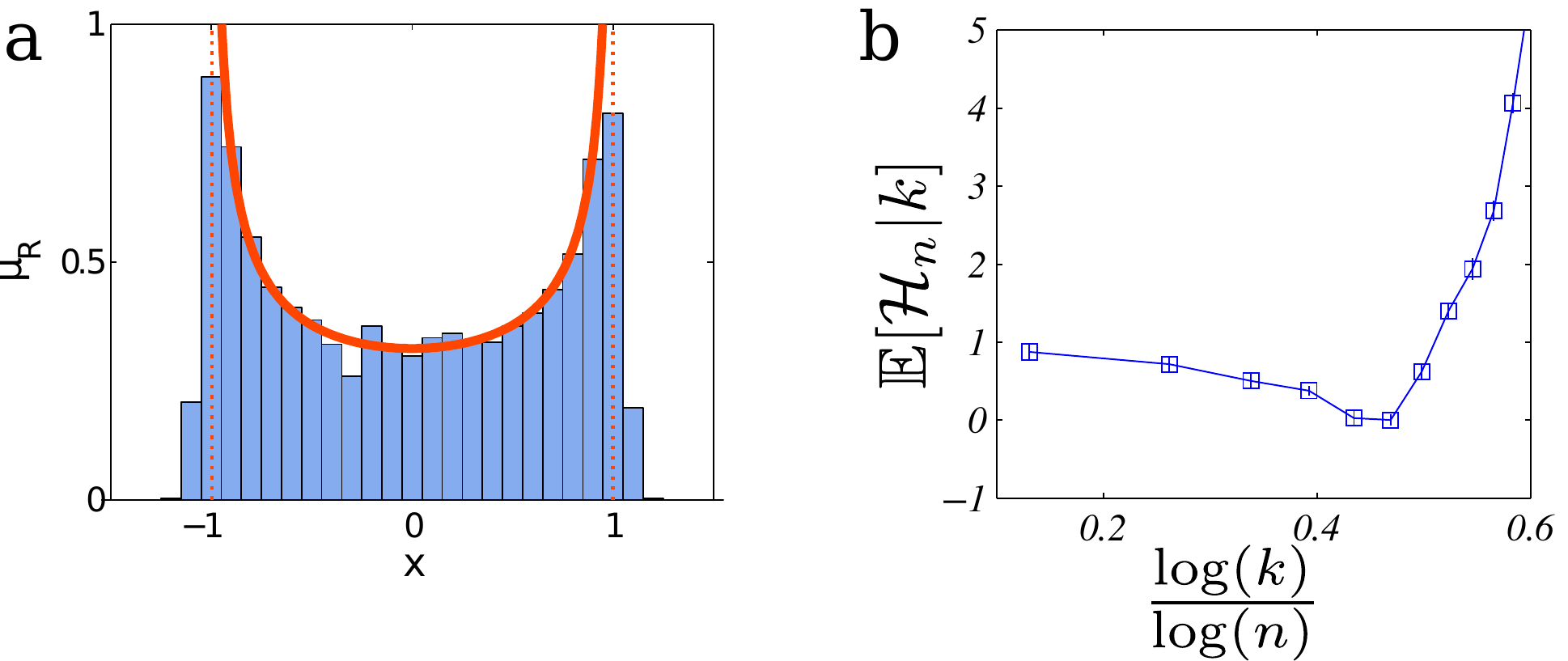}
 \caption{ \small  \textbf{(a)} $\mu_R$ for the Ginibre ensemble with $n=200$ and $k=2$ averaged over $5000$ realizations with the inverse semi-circular law $\mr(x)=(\pi \sqrt{1-x^2})^{-1}$ superposed. \textbf{(b)} Average energy of constrained matrices of size $n=200$, the error bars are the size of the markers. There is a clear minimum at $k=\O{\sqrt{n}}$. }\label{fig:twofigures}%
\end{figure}

Now that we have characterized the statistics of $k_n$, we investigate their empirical distribution $\mr$ on the real line for fixed $k_n$. We observe that the macroscopic behavior of $\mr$ undergoes an outstanding transition between a unimodal distribution for $k_n\gg\bar k_n$ and a bimodal distribution for  $k_n\ll\bar k_n$. The log-gas analogy reveals that this transition is related to a fine competition between the confinement $W$ and the forces exerted by eigenvalues near the real axis (see Methods). Around real eigenvalues, a \emph{gap} of size $\zeta(k_n)$ is formed, due to the effects of mirror repulsion between complex conjugated eigenvalues and repulsion of real eigenvalues. This gap modifies the effective potential on the real axis making it flatter as $k_{n}$ decreases; confined real eigenvalues concentrate at $0$ (unimodal distribution) while real eigenvalues on a sufficiently flat potential escape the center (bimodal distribution) converging to the \emph{inverse semi-circular law} in the limit of very small $k_n$ (see figure \ref{fig:twofigures} a).

The arguments provided before show that the distribution of real eigenvalues interpolates between that of strongly confined systems for $k_n=\O{n}$ and weakly confined systems for $k_n=o(n)$ and therefore, the spacings distribution $f(s)$ interpolates between the Wigner surmise and semi-Poisson statistics (see figure~\ref{fig:main}). We also remark that the distribution of extremal eigenvalues interpolates between Tracy-Widom~\cite{tracy1994level,tracy1996orthogonal} and Rider-Sinclair\cite{rider2014extremal} and the correlations between real eigenvalues interpolate from those of determinantal processes \cite{dyson62b} and those of Pfaffian processes \cite{Forrester07} (see Supp. Mat.).

\begin{figure}[t]
\centering
\includegraphics[width=.48\textwidth]{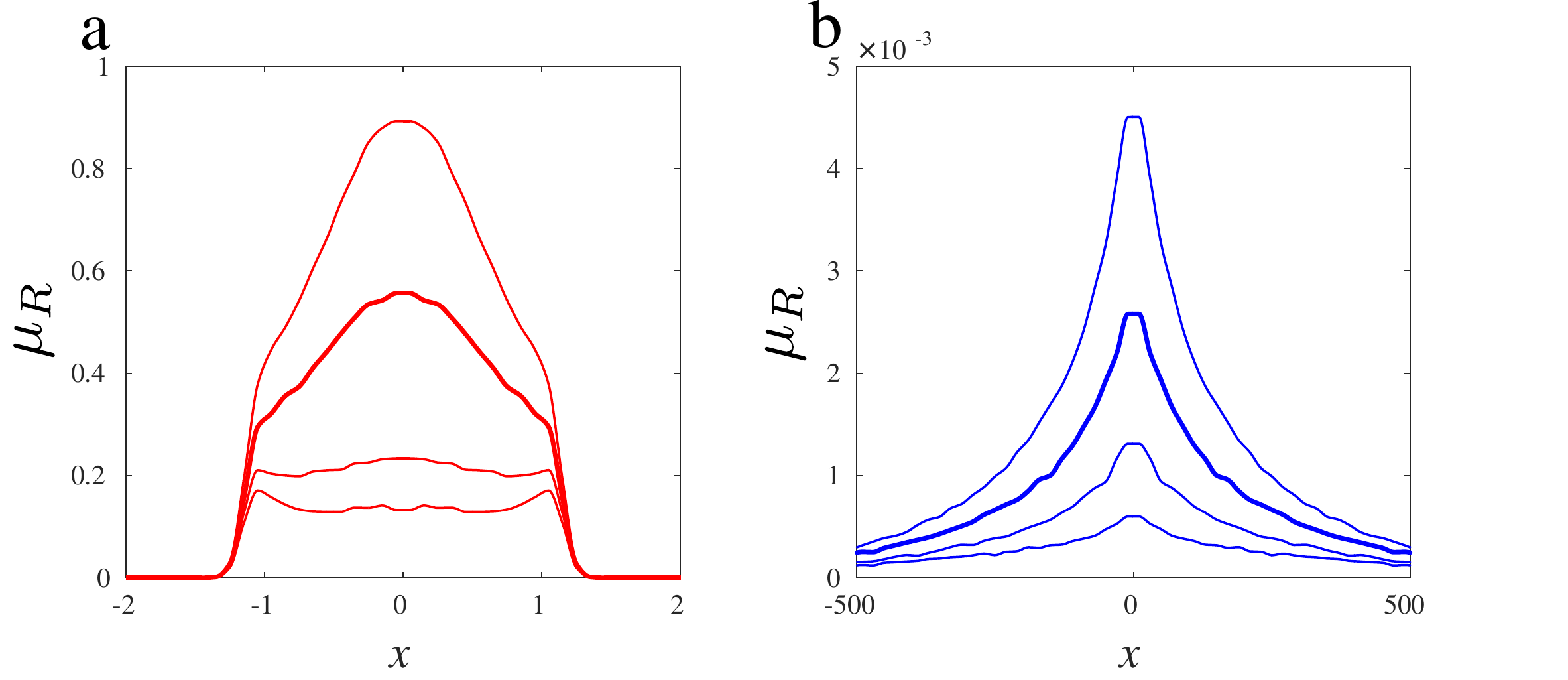}
 \caption{ \small  \textbf{Transition of $\mu_R$ for non \iid matrices. (a)} Transition in $\mu_R$ for \im matrix 1 with $\theta=0.3$ and $n=200$. Each curve corresponds to a different value of $k_n$,  the thick line corresponds to the most likely value of $k_n$ (closest to $\bar k_n$). Histograms have an area $k_n/\bar k_n$. \textbf{(b)} Absence of transition in $\mu_R$ for \h stable matrix with $\alpha=1$ and $n=200$. (see Methods for the details about the construction of the matrices.) }\label{fig:transition_i_stable}
\end{figure}

Our numerical simulations confirm these results and moreover, show that this transition is also present in all matrices analysed with independent identically distributed entries and bounded second moment. A similar transition arises for matrices with independent but non identically distributed entries (see figure \ref{fig:transition_i_stable}). In contrast, the distribution $\mr$ for the class of matrices with unbounded second moments did not show a significant dependence on the value of $k_n$.

The estimate of the size of the gap $\zeta(k_n)$ can also account for the order of magnitude $k_n=\O{\sqrt{n}}$ as a minimizer of the energy (see figure~\ref{fig:twofigures} b). Indeed, for $k_n=\o{\sqrt{n}}$, placing a pair of non-real eigenvalues on the real axis decreases the energy by reducing the packing of non-real eigenvalues leaving unchanged the size of the gap. When $k$ exceeds $\O{\sqrt{n}}$, this displacement requires the energy of increasing the size of the gap. 

We have thus exhibited two new properties related to the number and distribution of real eigenvalues $k_{n}$. First, we showed that the number of eigenvalues of the Ginibre ensemble is asymptotically Gaussian, and that this property is valid for a wide class of matrices. This result is evocative of a number of central limit theorems for the number of elements of determinantal processes in disks~\cite{costin1995,soshnikov2000,soshnikov2002}, although real eigenvalues are a Pfaffian process~\cite{Forrester07} confined on a singular region of the complex plane. This is probably a reason for the discrepancy in the level of fluctuations in both cases: $\O{n^{1/8}}$ for the complex Ginibre ensemble in a ball of radius $1/\sqrt{n}$~\cite{Tao10}, substantially smaller than fluctuations of the number of real eigenvalues $\O{n^{1/4}}$, which may be related to the fact that only outermost eigenvalues contribute to the fluctuations in a ball whereas all real eigenvalues contribute to the fluctuations of $k_n$.

We have also exhibited a transition in the distribution of real eigenvalues between a unimodal and a bimodal distribution~\footnote{So-called P-bifurcation in random dynamical systems~\cite{arnold1998random}}, with spacings interpolating between Wigner surmise and semi-Poisson statistics, when the scaling of $k_n$ is varied. Interestingly, this transition, which is already visible for finite $n$, ensures that even if the number of real eigenvalues is small, the probability that the eigenvalue with largest real part is real does not vanish. This result is of particular importance for stability properties and dynamics of randomly connected neural networks~\cite{garcia:13} (see Supp. Mat.). 

While valid for a wide class of matrices, this transition was not found for matrices with unbounded spectrum, which is also the case of log-gases within sub-quadratic potentials that do not compactly confine particles. These gases, like heavy-tailed random matrices show unimodal distributions of real particles $\mr$ at any scaling. This consistency suggests that the spectrum of large non-Ginibre matrices distribute as log-gases within specific potentials. 

All these findings go beyond the case of real eigenvalues, and are valid for the number and distribution of particles $h_{n}$ at a distance $l(n)=\o{1}$ of any given axis $\mathcal{A}$ and for the number of eigenvalues in elongated regions of the complex plane such as rectangles with a microscopic width $l(n)=\o{1}$ and a macroscopic height $L=\o{1}$. In both cases, the number of eigenvalues $h_{n}$ fluctuates asymptotically as a Gaussian with mean and variance
$\O{nl(n)}$, and the distribution along the axis $\mathcal{A}$ (or along the height of the rectangle) transitions from a bimodal shape for $h_{n}=\o{nl(n)}$ to a unimodal profile for $h_{n}=\O{n}$ through a uniform distribution (see Supplementary Material). The microscopic observations such as the presence of margelles also hold, and may be the counterpart of Dirac mass observed in macroscopic conditionings~\cite{Anderson09}. But a unique aspect of real eigenvalues is that the gap forms naturally from the mirror repulsion of eigenvalues, and thus these phenomena occur even when constraining on a singular region of the complex plane.

\bibliographystyle{apsrev4-1}  
\bibliography{biblio.bib}

\cleardoublepage

\appendix

\centerline{{\bf \textsc{Methods}}}

\section{Generating rare matrices}

In order to generate random matrices constrained to have a prescribed number of real eigenvalues $k_n^*$, we developed an algorithm based on rejection sampling and Monte-Carlo perturbations of single entries. Starting from a given random matrix with $k\neq k_n^*$ real eigenvalues, 
we iteratively redraw one of the entries chosen at random and accept this change if the new number of real eigenvalues $k'$ is closer to $k^*_n$ or equal to $k_n$ until reaching $k^*_n$. 


\section{Three classes of random matrices}
We have considered three main families of random matrices with increasing discrepancy with the Ginibre ensemble. We have chosen these families according to how their limiting empirical spectral distributions (ESDs) deviate from the circular law in order to shed light on the relation between the ESD and the statistics of real eigenvalues.

\begin{itemize}
	\item {\underline{\iid matrices}: this family of matrices corresponds to matrices with independent, identically distributed elements with bounded second moments. We used four distinct distributions in the simulations:} (i) Gaussian, (ii) centered uniform, (iii) centered Bernoulli and (iv) a distribution from the Pearson VII family with unbounded 4th moment ($f(x)=3(2+x^2)^{-5/2}$). These matrices belong to the universality class of the Real Ginibre ensemble in the sense that {their ESDs and local properties converge to the same limit}~\cite{Tao10,Tao12}. Despite the existing universality results for this family, a number of properties have never been analyzed, including for instance the distribution of real eigenvalues $\mr$. 
	\item {\underline{\im matrices}: this class corresponds to random matrices with independent but not identically distributed elements with bounded second moment. These matrices generally have ESDs with compact support that are distinct from the circular law. Specifically, we consider matrices} $D^LXD^R$ where $X$ is a Ginibre matrix, and $D^L$ and $D^R$ are diagonal matrices whose diagonal elements, given by the vectors $L$ and $R$, are assumed to be random and such that $\E[L_iR_j]=1$ for all $i$ and $j$. A particularly important quantity related to these matrices is given by $\theta=\lim \sqrt{\mbox{Var}(L) +\mbox{Var}(R)}$. We investigate two models: (i) \im matrix 1 correspond to $L=\1_n$ and $R$ is a Gaussian random vector, and (ii) \im matrix 2 correspond to $L$ and $R$ both being Gaussian random vectors \footnote{We expect that the same results hold for cases where $X$ does not belong to the Ginibre ensemble, but with independent elements having the same second moment, but might be drawn from different distributions}. The asymptotic ESD of a subclass of \im matrices were investigated in~\cite{Rajan06,ahmadian15,yiwei}. Nothing is known about real eigenvalues distribution for these matrices.
	\item {\underline{\h matrices}: this class correspond to matrices whose elements} have unbounded second moment, and with density $df(x)\sim |x|^{-1-\alpha}$ at infinity. These matrices have an unbounded spectrum. For the sake of simplicity we consider independent identically distributed entries. In particular we analyse the symmetric ($\beta=0$) and the skewed ($\beta=1$) $\alpha$-stable distributions (see e.g.~\cite{zolotarev1986one}) as well as the non stable distribution $f(x)\propto|x|^{-1-\alpha}\mathbbm{1}_{|x|\geq 1}$ for $\alpha=\{0.5,1\}$. The asymptotic ESDs, and particularly their tails, were investigated  in~\cite{bordenave11}. The statistics of real eigenvalues and their distribution were not studied.
\end{itemize}
In figure \ref{fig:distributions} (Supp. Mat.) we provide numerical evaluations of the ESD and distribution of real eigenvalues for these matrices. 

We also consider the generalized eigenvalues of two Ginibre matrices which is the same as the spectrum of the quotient of two Ginibre matrices. These matrices were introduced in \cite{Edelman94} and coined Cauchy matrices. It is known that the number of real generalized eigenvalues scales as $\sqrt{\pi n/2}$ and $\mu_R=1/\pi(1+x^2)$. These are related to \h matrices because their spectrum is also unbounded but they are different from all the examples considered before because their entries are not independently distributed. However, the correlations decrease to $0$ as the matrix size increases.

A crossover family between \im matrices and \h matrices can be constructed by having non identically distributed entries with unbounded second moment. In such cases we expect the statistics to be continuous in the sense that they converge to the limits presented before as the heterogeneity decreases or as the second moments get bounded.

\section{Analytical justification of the Gaussian fluctuations of $k_n$ for Ginibre matrices}

In this section, we provide analytical developments to justify the Gaussian nature of the distribution of $k_n$ for large Ginibre matrices.
Our approach is based on the log gas analogy for the distribution of eigenvalues of Ginibre matrices. This analogy is based on the identity between joint probability distribution of eigenvalues with a prescribed number of real eigenvalues, given in closed form in~\cite{Lehmann91,Edelman97} and the distributions of particles in a specific Coulomb gas. Similarly to the Gaussian ensembles and complex Ginibre ensemble~\cite{Dyson62,Forrester10}, we observed in~\cite{garcia-JSP:16} that the joint distribution of the eigenvalues $(\lambda_{1}\cdots \lambda_{n})$ is identical to the Gibbs measure of a two phase log-gas in a convex potential $W$ given explicitly in~\cite{Dyson62,majumdar2009index,majumdar2011many,Forrester10}, and at a inverse temperature $\beta=1$. 

In this gas, the typical distance between particles is of order $1/\sqrt{n}$, which is exactly the order of magnitude of the noise, hence particles that are a few typical distances away have extremely weak correlations and in particular, eigenvalues separated by an asymptotically infinite number of eigenvalues are independent. With this decorrelation distance in mind, we divide the interval $[-1, 1]$ into $p(n)$ disjoint subintervals $\{I_i=[a_i,b_i]\}_{1\leq i\leq p(n)}$  of length $|I_i|=c(n)$ and such that $p(n)\to\infty$ and $c(n)\sqrt{n}\to\infty$. These intervals are separated by a distance $\delta(n)$ satisfying $\delta(n)\sqrt{n}\to\infty$ and $\delta(n)/c(n)\to0$. 
 We denote $\kappa_i$ the number of eigenvalues in $I_i$ which given our construction, diverge in average and are asymptotically independent because
 two consecutive intervals are separated by a diverging number of eigenvalues. Since $\delta(n)/c(n)\to 0$, we obtain at leading order:\vspace{-2mm}
 \[\frac {k_n}{\sqrt{n}}=\sum_{i=1}^{p(n)}\frac{\kappa_i}{\sqrt{n}} + \O{\frac{p(n)}{\sqrt{n}}}.  \]
This expression of $k_n$ as sum of independent random variables ensures, by virtue of the Central Limit Theorem, that its fluctuations are Gaussian:
 \[k^\star_n=\frac{k_n-\bar k_n}{\sqrt{(2-\sqrt{2})\bar k_n}}\to\mathcal{N}(0,1)\ .\]
This approach actually allows uncovering the fine structure of the fluctuations of real eigenvalues. We can indeed generalize this calculation and show that the number of real eigenvalues $k_n(I)$ within an subset $I\subset [-1,1]$ of size $
\vert I\vert=\O{1}$ is such that 
\[k_n^{\star}(I) = \frac{k_n(I)-\sqrt{\frac{n}{2\pi}} \vert I \vert}{\sqrt{(2-\sqrt{2})\sqrt{\frac{n}{2\pi}} \vert I \vert}}\to\mathcal{N}(0,1)\ .\]
Moreover, for any two intervals $I_1$, $I_2$ separated by a distance $d(I_1,I_2) \sqrt{n}\to \infty$, we have $(k_n^{\star}(I_1),k_n^{\star}(I_2))\rightarrow \mathcal{N}(0,\1_2)$ where $\1_2$ is the identity matrix in dimension $2$. This implies in particular that $k_n^{\star}([-1,x])$ has Gaussian independent increments and therefore is the Brownian motion on $[-1,1]$.

These results can be shown to be universal based on the powerful the universality theorem~\cite[Theorem 22]{Tao10}. Indeed, applying this theorem would indeed allow to show that all moments of $k_n^\star$ (or $k_n^{\star}(I)$ for $I\subset [-1,1]$) converge towards those of the standard Gaussian variable for any matrix in the universality class of~\cite{Tao10}, thus implying the universality of the Gaussian nature of fluctuations in this class~\footnote{Note that rigorously, a complete proof of this result shall use the fact that the convergence of the moments is uniform on bounded regions~\cite[Theorem 12]{Tao12}, implying a uniform convergence of the characteristic function of $k_n$ for matrices in this universality class, hence convergence in law of $k_n^\star(I)$ towards the standard Gaussian variable.}.

In the main text, we have shown that the universality of this result goes way beyond this class. This statement was based on extensive simulations of matrices, and was statistically validated using the two-sample Kolmogorov-Smirnov 
test comparing $p^n_k$ to the distribution of a discretized Gaussian variable obtained by rounding to the closest even integer a Gaussian 
with mean $\bar k_n$ and variance $\Sigma^2_n$.

\section{Transition in the distribution of $\mu_R$}

In this section, we provide analytical arguments, based on the log-gas analogy, to account for the transition in $\mu_R$ for Ginibre matrices. Our arguments are based on monitoring precisely the effective potential felt by real eigenvalues (produced by both confinement and non-real eigenvalues).
 
We first consider the distribution of non-real eigenvalues as a function of the order of magnitude of $k_{n}$. For $k_n=\O n$, the real eigenvalues induce a macroscopic repulsion that pushes non-real eigenvalues at a strictly positive macroscopic distance of the real axis (the ESD of such matrices thus does not converge towards the circular law, see~\cite{garcia-JSP:16}). As soon as $k_n=o(n)$, the empirical spectral distribution $\mu$ converges to the circular law $\mu_0$. At the microscopic scale, a gap forms of typical size $\zeta(k_n)=\o{1}$ around the real axis, resulting of both mirror repulsion between conjugated eigenvalues and the repulsion generated by real eigenvalues. The respective role of mirror repulsion and real eigenvalues repulsion controls the size of the gap. When $k_n=o(\sqrt{n})$ the mirror repulsion dominates and $\zeta(k_n)$ is of the order of the distance between complex particles $\O{1/\sqrt{n}}$. If $k_n=\O{\sqrt{n}}$ or larger, the repulsion generated by the real gas 
dominates and since the leading term of the confining potential is quadratic and the interactions are electrostatic, $\zeta(k_n)$ scales as $k_n/n$. 
These estimates allow us to demonstrate the transitions in the shape of $\mr$ as the scaling of $k_n$ is varied. 

In order to handle rigorously this question, one can express the empirical spectral distribution as a perturbation of the circular law: $\mu=\mu_0+\mu_R+\mu_{C}$, where $\mu_{R}$ is a positive measure on the real line with mass $k_n/n$ accounting for the presence of real eigenvalues, and the signed measure $\mu_C$ on the complex plane with mass $-k_n/n$ introduces a gap on the region $S_\zeta=\{z,|\Im(z)|<\zeta,|\Re(z)|<1\}$ and the mass removed from the gap is denoted $\zeta G$. When the mass removed from $S_{\zeta}$ is larger than $k_n/n$, the excess $\gamma=(\zeta G-k_n/n)$ is relocated inside the unit disc $D=\{z,|z|\leq1\}$. We write the energy of the gas as 
\begin{align*}
 \mathcal{H}(\mu)=
 &K+\mathcal{H}_{out}(\mu)- \frac{n^2}{2}\iint\log|x-x'|(d\mu_R+d\mu_{C})^2 
\end{align*}
where $K=\H(\mu_0)$ is a constant and $\H_{out}(\mu)= n\int W(x)\1_{|x|>1}d\mu(x)$ is the energy corresponding to the eigenvalues with $|\lambda|>1$ (the confining potential is essentially compensated by the circular law inside $D$ but not outside). This term prevents the eigenvalues from leaving $D$ as long as $k_n=o(n)$.

Therefore we distinguish three limit regimes depending on the scaling of $k_n$ (see Fig.~\ref{fig:main}): First, in the saturated regime ($k_n/n\to\alpha$ with $0<\alpha<1$) the presence of a macroscopic gap reduces the influence of complex eigenvalues on the real gas. For this reason, the interaction among eigenvalues does not compensate the quadratic confining potential over the real axis, and thus the shape of $\mr$ becomes concave. In that regime, the support of $\mr$ exceeds the unit interval $[-1,1]$. Indeed, the term $H_{out}$ penalizing real eigenvalues outside the unit interval is now of the same order of magnitude as the other terms of the energy, and therefore a trade off will be found leading to a support strictly exceeding $[-1,1]$. This is perfectly consistent with the mathematical developments and abstract characterizations provided in~\cite{garcia-JSP:16}.

Second, in the central regime ($\O{\sqrt{n}}\leq k_n \leq o(n)$) the empirical spectral distribution converges towards $\mu_0$. The gap becomes microscopic and can accommodate the possible excess or lack of real eigenvalues: its size adjusts so that $\zeta(k_n)=\frac {k_n}{nG}$, ensuring that $\gamma=0$. The perturbation of the energy generated by $\mu_C$ is completely absorbed by adjusting the gap and real eigenvalues minimize their energy when $\mr$ is uniform in $[-1,1]$. In finite size matrices the transition between the central and saturated regimes is progressive through a dense regime.

Third, for the diluted regime ($k_n= o(\sqrt{n})$) the gap can no longer accommodate the excess of eigenvalues removed from the real line because its size $\zeta(k_n)$ remains at least of order $\sqrt{n}$. In that case, we see that $\gamma$ becomes strictly positive for all finite $n$. An excess of eigenvalues accumulate outside the gap, forming microscopic bumps (the \emph{margelles}). In this regime, the concave potential created by the non-real eigenvalues becomes stronger, and as a result the potential felt on the real axis becomes too flat to be compensated by a uniform distribution on the real axis: the eigenvalues are pushed away from $0$ and organize in a bimodal distribution. 

Finally, let us provide an explicit form for the distribution of real eigenvalues in the limit of very small $k_n$. In that case, the variational description of $\mr$ as a minimizer of the energy provides for any $x\in[-1,1]$ the integral equation:
\[{\int_{-1}^{1} \frac{1}{x-x'}d\mr(x') = 0}\] 
subject to the normalization and positivity constraints. This equation characterizes the distribution of electrostatic particles in a flat-well potential. Actually, this very equation was previously studied for its applications in aerodynamics, and the solution is provided in~\cite[Chapter 4.3]{tricomi}: 
\[{\mr^0(x)=(\pi \sqrt{1-x^2})^{-1}}.\] 
We thus conclude that distribution of real eigenvalues, in the limit of very small $k_{n}$, converges towards \emph{inverse semi-circular law}. Numerical simulations show an excellent agreement with this theory, see Figure~\ref{fig:twofigures} a.

\cleardoublepage

\centerline{{\bf \textsc{Supplementary material}}}

\section{Statistics of the number of real eigenvalues in matrices with strongly correlated entries}

In the main text we have shown that the Gaussian nature of the fluctuations of the number of real eigenvalues was valid for a wide range of real random matrices with independently distributed entries. We have found in all cases that the variance of the number of real eigenvalues satisfied the relationship $\Sigma_{n}^{2}=(2-\sqrt{2})\bar{k}_{n}$. We provide here a few examples of matrices with correlated coefficients for which this relationship also not hold, despite the fact that the scaling of $\bar k_n$ can take arbitrary values. These include partially symmetric and almost symmetric matrices, or products or random matrices, that we discuss in the following subsections.

\subsection{Partially symmetric and almost symmetric matrices}
First we consider matrices $(m_{ij})_{1\leq i,j \leq n}$ whose entries satisfy $\E[m_{ij}]=0$, $\E[m_{ij}^2]=1$ and $\E[m_{ij}m_{kl}]=\delta_{il}\delta_{kj}\tau$ for $i\neq j$ with $-1\leq\tau\leq 1$. At $\tau=1$, the matrices are symmetric and hence $k_n=n$. Conversely, for $\tau=-1$, $k=0$ (or $1$ if $n$ is odd). In \cite{sommers88} the limiting ESD was given. For $-1<\tau<1$ the ESD is uniform inside the ellipse centered at 0 whose axis are over the real and imaginary axis and have length $2 + 2\tau$ and $2-2\tau$ respectively. The joint  eigenvalues probability distribution for fixed $k$ and $\tau$ was given in \cite{Lehmann91}. Despite these remarkable works, the number or the distribution of their real eigenvalues has never been studied.

To study the limiting behaviour of the real eigenvalues we define a sequence $\epsilon_n\to0$ and we distinguish two limits: partially symmetric (antisymmetric) matrices for which $-1+\epsilon_n<\tau<1-\epsilon_n$ and almost symmetric (antisymmetric) matrices for which $\tau_n>1-\epsilon_n$ ($\tau_n<-1+\epsilon_n$).

Since the joint probability distribution for a fixed $k$ and $1>\tau>-1$ is known, a log-gas approach can be derived. Therefore, the results shown above for Ginibre matrices also apply to these matrices. For any $\tau_n\to\tau^*$ there is a limit $\bar k(\tau^*)/\sqrt{n}\to(1+\tau^*)\sqrt{2/\pi}$, which increases as the length of the horizontal axis of the ellipse increases. Also $\Sigma^2=(2-\sqrt{2})\bar k$ and the fluctuations around the mean are Gaussian. 

\begin{figure}[h]
\centering
\includegraphics[width=0.5\textwidth]{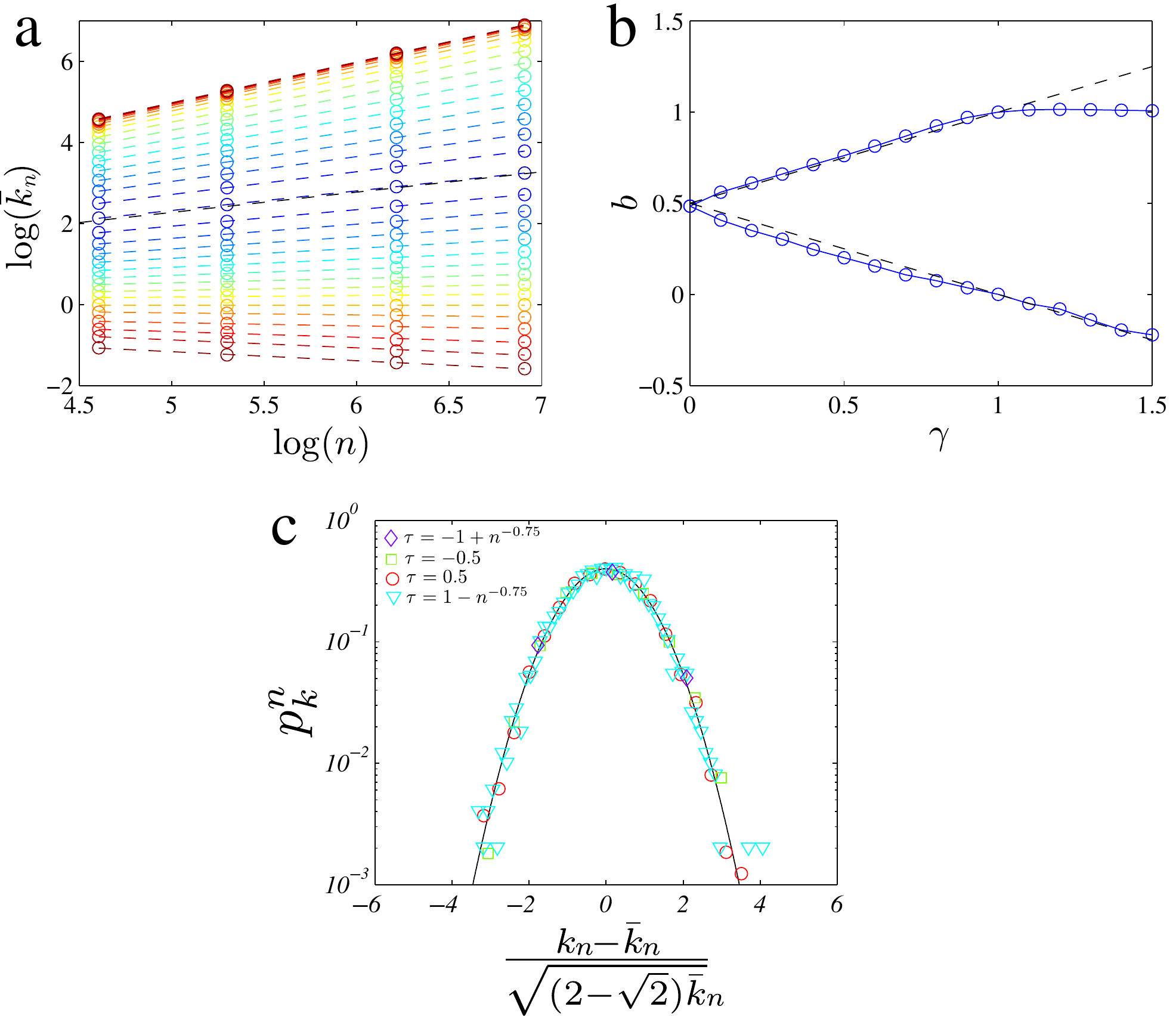}\\
\caption{\small \textbf{Statistics of $k_n$ for partially and almost symmetric matrices. (a)} Numerical estimations of $\bar k$ for almost symmetric matrices of different sizes and with $\gamma=\{0,0.1,0.2,\dots,1.5\}$ (blue corresponds to $\gamma=0$, red  to $\gamma=1.5$). The dashed lines are linear fits. \textbf{(b)} Scaling of $\bar k$ (parameter $b$) as a function of $\gamma$ for the two limits $\tau\to\pm 1$. The dashed lines have slopes $\pm 0.5$. The upper branch saturates at $b=1$ (i.e.~$k=\O{n}$). \textbf{(c)} Fluctuations of $k_n$ for partially and almost symmetric matrices. Desptite having a differnt scaling for $\bar k_n$, these matrices have the same ratio $\Sigma^2/\bar k_n$ and $k_n$ has Gaussian fluctuations. Here $n=1000$. }\label{fig:k_of_tau_of_n}
\end{figure}

The case where $\tau_n\to1$ and $\tau_n\to-1$ is more subtle. Indeed, one can find a sequence $\tau_n$ such that $\E[k_n]$ has any given scaling. In order to have a scaling for $k$ different from $\O{\sqrt{n}}$, $\tau_n$ has to converge to $1$ or $-1$ as $n\to\infty$ and the scaling will depend on the speed of convergence of $\tau_n$. In our simulations we study the scaling of $\bar{k}_n$ in two limits: 
\begin{align*}
\tau_n&=1-n^{-\gamma}\\
\tau_n&=-1+n^{-\gamma}\ .
\end{align*}
Figure \ref{fig:k_of_tau_of_n} shows numerical results for these scalings. First we see that for each limit the behaviour of the average number of real eigenvalues is
\[\bar{k}_n=an^b\] 
for constants $a$ and $b$ that depend monotonically on $\gamma$. One can also see that the two limits are different. In the limit $\tau\to1$ $b$ increases with $\gamma$ and the contrary happens in the limit $\tau\to-1$, however $b$ is upperbounded by $1$ and instead it has no lower bound. Furthermore, our simulations show that for fixed $n$ the convergence of $\E[k_n]$ to its limits as $\tau\to1$ ($\tau\to-1$) is non-decreasing (non-increasing).

\subsection{Products of real random matrices}
\begin{figure}[thb]
\centering
\includegraphics[width=0.5\textwidth]{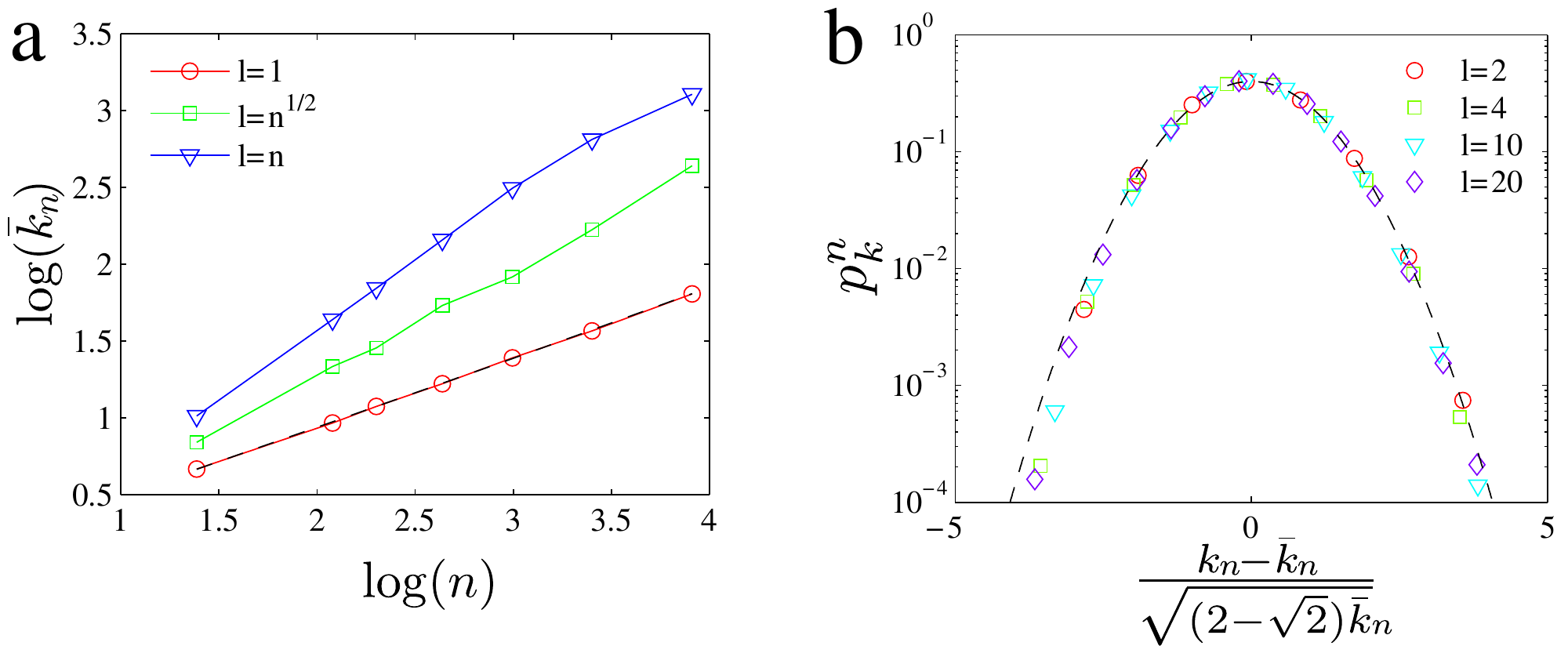}
\caption{\textbf{Statistics of $k_n$ for products of matrices. (a)} Scaling of the average number of real eigenvalues of products of Ginibre matrices for different values of $l$. The case $l=1$ matches perfectly the expansion in \cite{Edelman94} (dashed line) as expected and $\bar{k_n}=\O{\sqrt{n}}$. For $l=\O{\sqrt{n}}$ and $l=\O{n}$ the scaling of $\bar{k_n}$ is larger than $\sqrt{n}$. \textbf{(b)} $p_k^n$ for rescaled $k$ with $n=50$ and several values of $l$. Again, for these matrices also  have the same ratio $\Sigma^2/\bar k_n$ and $k_n$ has Gaussian fluctuations.}\label{fig:k_of_K}
\end{figure}

Second we consider products of real \iid matrices. The product of complex Ginibre matrices has been analysed in \cite{burda10,burda10b}, however, the real case is very different because of the presence of real eigenvalues. In particular, denoting $M^l$ the product of $l$ real Ginibre matrices, for any matrix size $n$, in the limit $l\to\infty$, $k_n=n$ \cite{lakshminarayan13,forrester14} and $\E{k_n}$ is non decreasing for increasing $l$. Similarly to the previous case, we claim that one can find a diverging sequence $l_n$ such that $\E[k_n]$ has any given scaling larger than $\O{\sqrt{n}}$ (see figure \ref{fig:k_of_K}).

\section{Time correlations for the real eigenvalues of matrix-valued stochastic processes}
\begin{figure}[h]
\includegraphics[width=0.5\textwidth]{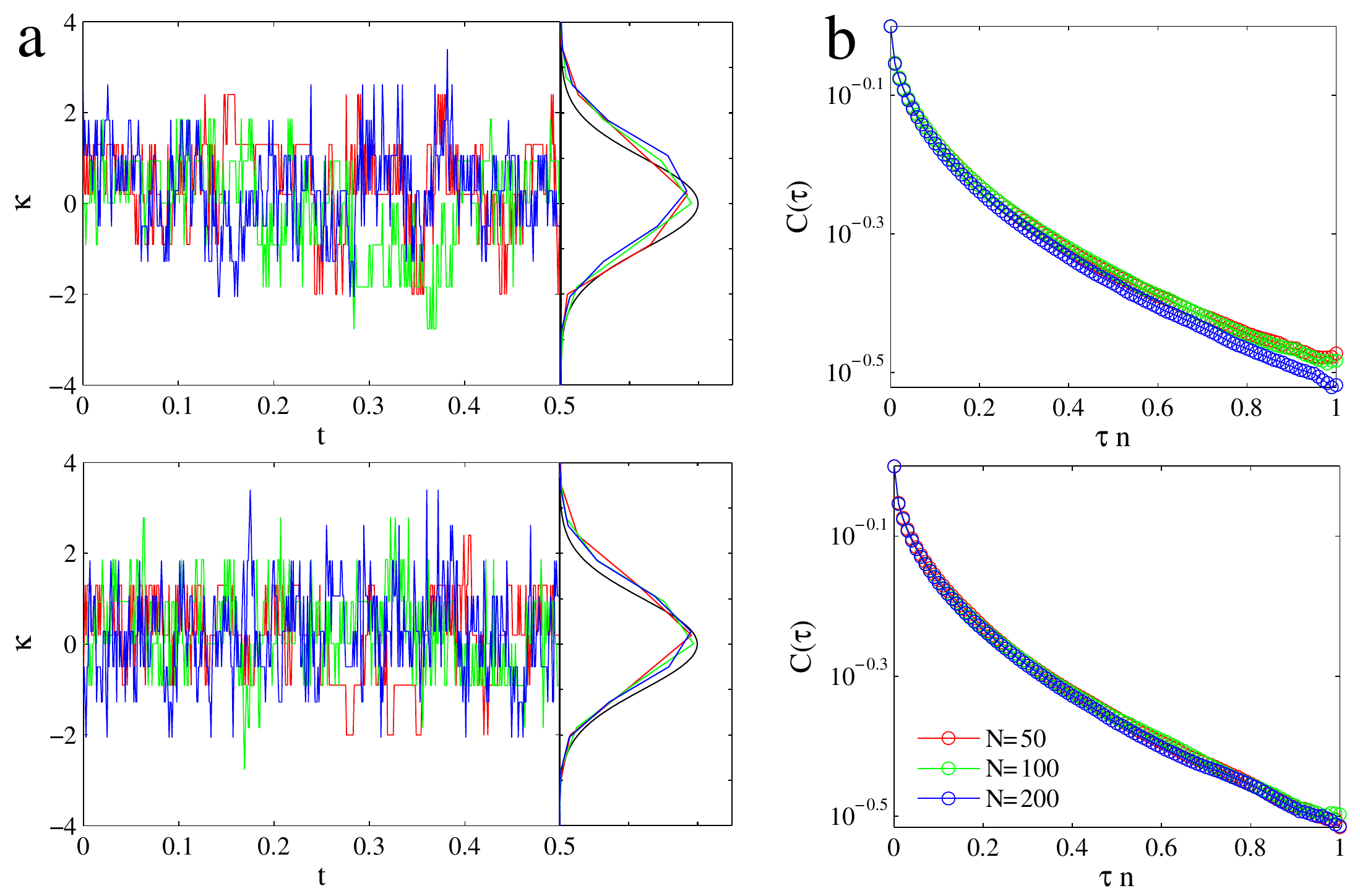}
\caption{\textbf{Real eigenvalues of matrix-valued stochastic processes. (a)} Trajectories of $\kappa(t)$ for different matrix sizes for the Wiener process (top) and the O.U. process (bottom). One can see that as $n$ increases the jumps of $\kappa(t)$ become more frequent. We also show the stationary distribution, i.e. the distribution of $\kappa(t)$ after a short transient (typically of order $1/n$) and in black the normal distribution. \textbf{(b)} The time correlations $C(\tau)$ for both processes decay exponentially, which is a signature of the brownian motion. Furthermore, correlations colapse to the same curve when the time axis is rescaled $\tau\to\tau n$. }\label{fig:k_of_t}
\end{figure}

\begin{figure}[h]
\includegraphics[width=0.5\textwidth]{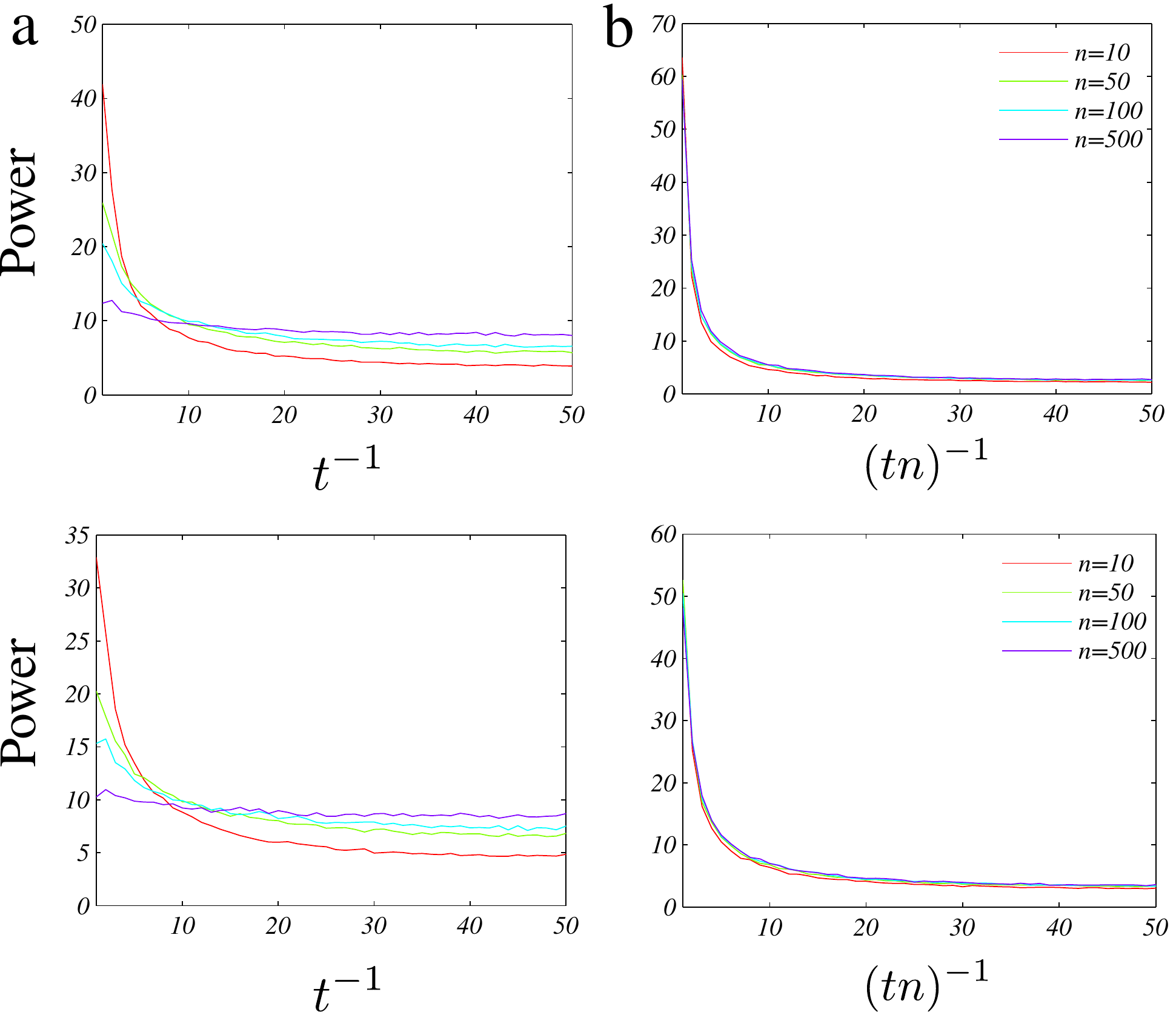}
\caption{\textbf{Fourier analysis of $k_n(t)$. (a)} Fourier spectrum of $k(t)$ for different matrix sizes for the Wiener process (top) and the O.U. process (bottom). As $n$ increases the the power spectrum becomes flatter converging to white noise. \textbf{(b)} Power spectrum for both processes as time is rescaled $t\to t n$. With that scaling the processes converge to a unique limit. }\label{fig:fourier}
\end{figure}

In relationship with the analysis of the statistics of $k_n$, we have mentioned that the Gaussian nature remained true when considering the number of real eigenvalues $k_n(t)$ for matrix valued stochastic processes. We provide here the details on the models analysed and the statistics of $k_n(t)$. First we consider and $n\times n$ Wiener process
\begin{equation}\label{eq:W}
dM_n(t)=dB_n(t),\quad M_n(0)=M_n^0
\end{equation}
where $B_n(t)$ is an $n\times n$ array of independent Brownian motions. Second we consider an $n\times n$ Orstein-Uhlembeck process:
\begin{equation}\label{eq:OU}
dM_n(t)=-M_n(t)dt+dB_n(t),\quad M_n(0)=M_n^0\ .
\end{equation}
We define the normalized quantity
\[\kappa(t)=\frac{k_n(t)-\bar k_n}{(2-\sqrt{2})\bar k_n}\]
and for a given initial condition $M_n^0$ we study the distribution of $\kappa(t)$ at different times and show that it quickly converges to a Gaussian. Furthermore we analyse its time correlations
\[C(\tau)=\mathbb{E}[\kappa(t)\kappa(t+\tau)]\]
where the average is done over different initial conditions $M^0_n$ (see figure \ref{fig:k_of_t}). Our simulations show that $C(\tau)$ decays exponentially with a characteristic time inversely proportional to the matrix size and hence $\kappa(t)$ converges to Gaussian white noise as $n\to\infty$ which we corroborate by estimating the Fourier spectrum of $\kappa(t)$ (see figure \ref{fig:fourier}).

This phenomenon can be understood geometrically. Indeed, a normalized $n\times n$ matrix can be represented as a point on the surface of the unit sphere in dimension $n^2$, and therefore an ensemble of random matrices is analogous to a density distribution over this sphere. In particular, the Ginibre ensemble corresponds to the uniform distribution. Since eigenvalues are a continuous function of the entries of the matrix, we can divide this surface into multiple regions according to the value of $k_n$ and neighbouring regions will have consecutive values of $k_n$. The processes given by \eqref{eq:W} and \eqref{eq:OU} when normalized are Brownian motions on the sphere. The spatial correlations will decrease exponentially and therefore so will the correlations in $k_n$.

\section{Extremal eigenvalues and spacing distribution}

{The main results have focused on statistical properties of eigenvalues. In this section we focus on extremal eigenvalues, specifically on the distribution of eigenvalues with largest real part that play a particular role in the stability of randomly connected networks~\cite{garcia:13}. Extremal eigenvalues of the real Ginibre ensemble were finely investigated for unconstrained matrices in~\cite{rider2014extremal}. The distribution of extremal eigenvalues for matrices with a prescribed number of real eigenvalues is still unknown. The results of the main text ensure that for $k_{n}=\O{n}$, the leading eigenvalue is real with probability 1. However, for $k_{n}=\o{n}$, this is no more necessarily the case, and the transition found in the shape of $\mr$ raises a profound question, since as the number of real eigenvalues decreases, they have a tendency to accumulate at the boundaries of the distribution.}

\begin{figure}[tbh]\centering
\includegraphics[width=0.5\textwidth]{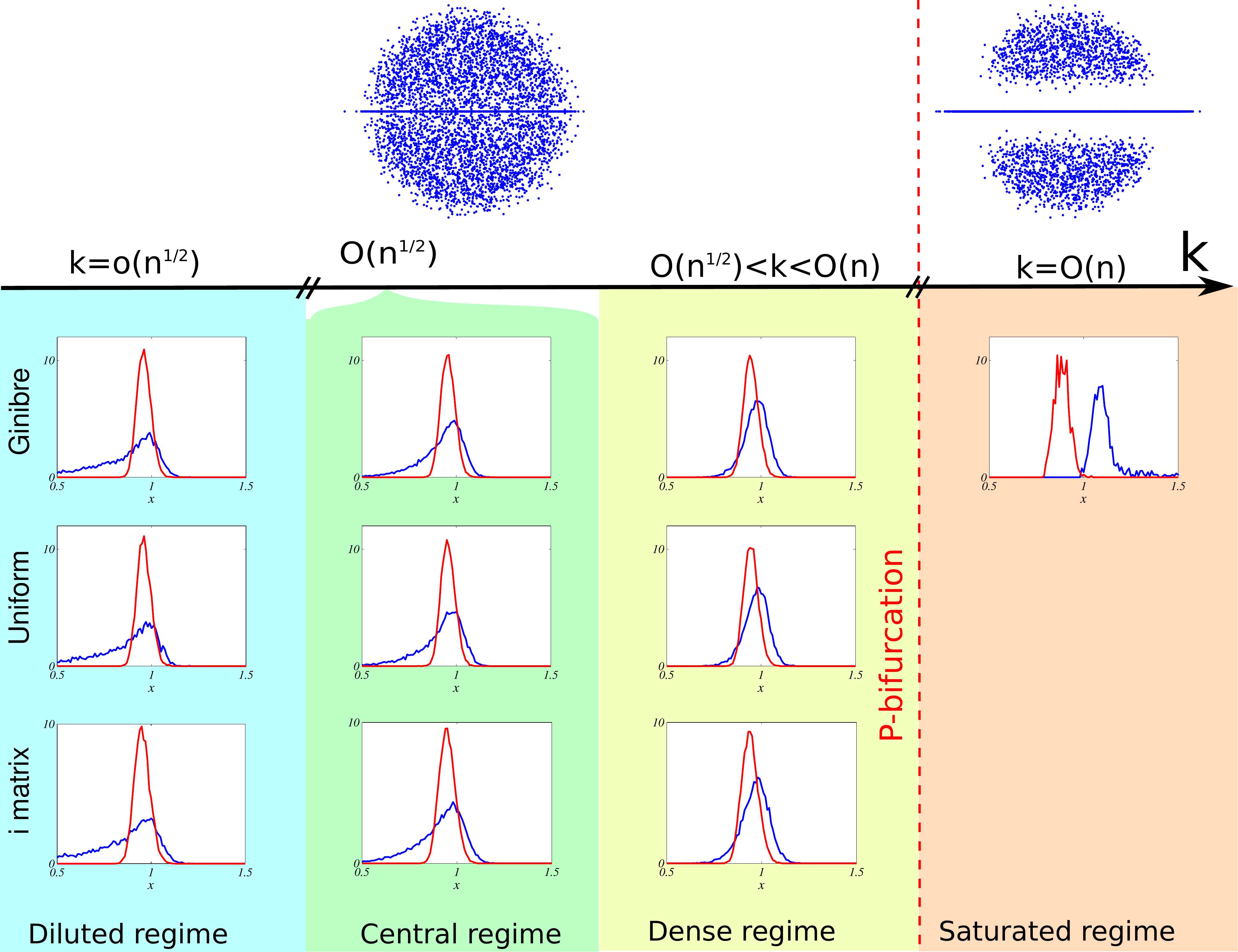}
\caption{\textbf{Extremal eigenvalues.} Distribution of the real eigenvalue with maximal real part (blue) and complex eigenvalue with maximal real part (red) for Ginibre, Uniform and \i matrices of size $n=200$ in the different regimes. In the saturated regime only gas simulations for Ginibre are accessible.}\label{fig:extremal}
\end{figure}

{In this section we investigate numerically the distribution of the largest real eigenvalue and of the complex eigenvalue with largest real part. }The key observations are summarized in figure \ref{fig:extremal}. The panels in this figure represent illustrative examples of the distributions of extremal real and complex eigenvalues for Ginibre, uniform and \im matrices matrices (upper, middle and lower rows, respectively) for different scalings of $k_{n}$. We observe that outside the saturated regime, i.e. for $k_n=o(n)$, the distribution of the complex eigenvalues with largest real part (red curves) hardly changes as $k_n$ varies, which is consistent with the fact that the asymptotic distribution of non-real eigenvalues in this regime remains unchanged. The situation is naturally very different regarding the largest real eigenvalues (blue curves). We observe that for small values of $k_n = \o{\sqrt{n}}$, the eigenvalue with largest real part has a broad distribution skewed towards small values. As $k_n$ is increased, the distribution becomes increasingly concentrated around $x=1$ and symmetric.
 
For $k_n=\O{n}$, the situation is markedly different. First, the support of the asymptotic distribution of non-real eigenvalues concentrates on a compact set included in $\{z\in\C; \vert Re({z})\vert\leq a<1\}$ (see~\cite{garcia-JSP:16}). Therefore, the distribution of the largest non-real eigenvalue shifts towards smaller values and is not centered at $1$. In contrast, the distribution of real eigenvalue exceeds the interval $[-1,1]$ and thus the eigenvalue with largest real part becomes peaked at value strictly larger than one. 

We now briefly discuss theoretically the origin of the distribution of spacings between eigenvalues. The distribution of eigenvalues of  the real Ginibre ensemble is proportional to the normalized mean spacing between consecutive real eigenvalues $s$, and therefore the distribution of spacings $f(s)$ goes to zero with finite slope at $s=0$. In the log-gas analogy, this is related to the logarithmic nature of the interactions between particles. The large $s$ behavior depends on the level of confinement. For $k_n=\O{n}$ the confinement is $\O{n}$ and as in the semicircular law results in a Wigner-surmise type distribution, with a tail that decays as $e^{-\pi s^2/4}$. For $k_n \ll n$ the confinement is very weak or non-existent and the tail of $f(s)$ corresponds to that of non confined particles (i.e. $f(s)\sim e^{-2s}$).

\section{Applications to the dynamics of randomly connected neural networks}
{The results on extremal eigenvalues presented in the previous supplementary section have a direct impact on the transition scenario for randomly connected neural networks studied in~\cite{garcia:13}. In this paper it is shown that a simple model of neuronal network with random balanced connectivity shows a transition as a function of the variance of the synaptic weights, either towards a stationary fixed-point regime when the eigenvalue with largest real part of the random connectivity matrix is real, or towards a synchronized oscillatory regime when that eigenvalue is non-real. The results of Figure~\ref{fig:extremal} allow to go deeper into the probability of each regime. 
In the case $k_{n}=\O{n}$, the transition to a stationary regime occurs with probability 1.}

\begin{figure}[H]\centering
\includegraphics[width=0.3\textwidth]{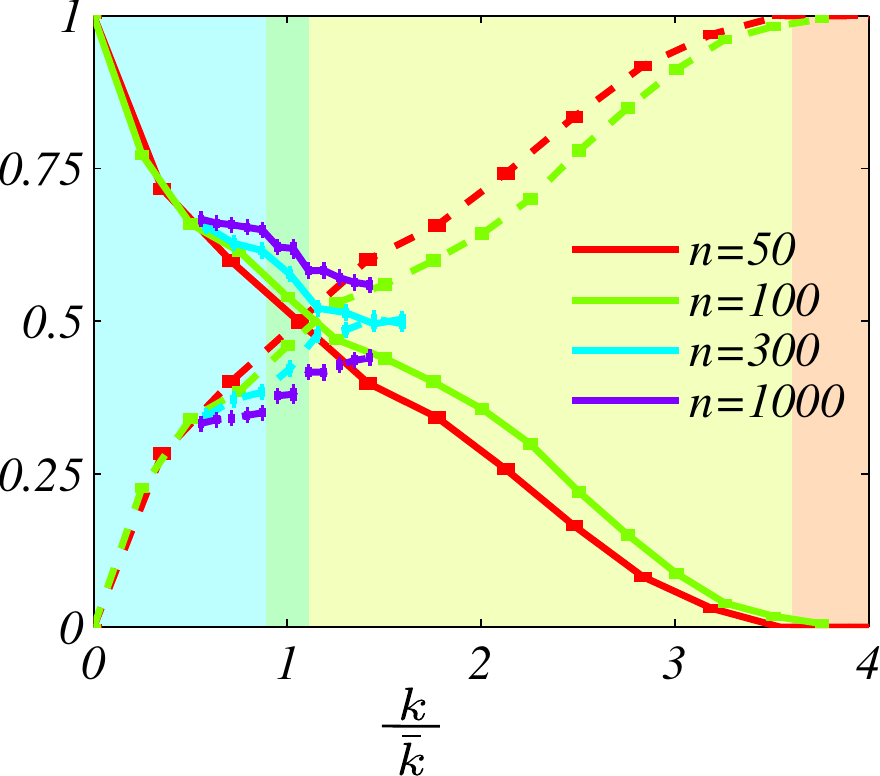}
\caption{Probability that the leading eigenvalues are complex (solid line) and real (dashed line) for Ginibre matrices of several sizes. The background colors represent the same regimes as in figure \ref{fig:extremal}. It is interesting to observe that the probabilities become similar when $k_n\ll\bar k_n$, i.e. in the saturated regime. Instead in the diluted and central regimes complex leading eigenvalues are more likely.}\label{fig:prop}
\end{figure}

However, in the case $k_{n}=\o{n}$, the situation is not clear since we observed that the distribution of real or non-real eigenvalues overlap in this regime. We provide in Figure~\ref{fig:prop} the probability of transitions to stationary regimes vs oscillatory regimes as a function of $k_{n}$ and $n$. We observe that when $k_n=\o{\sqrt{n}}$, although there are substantially less real eigenvalues, the bimodal nature of the distribution indicate that eigenvalues accumulate at the ends of the interval $[-1,1]$, possibly compensates for the reduced number of real eigenvalues in these regimes.

\section{Conditioning on other subsets of $\C$}
\begin{figure}[H]
\centering
\includegraphics[width=0.5\textwidth]{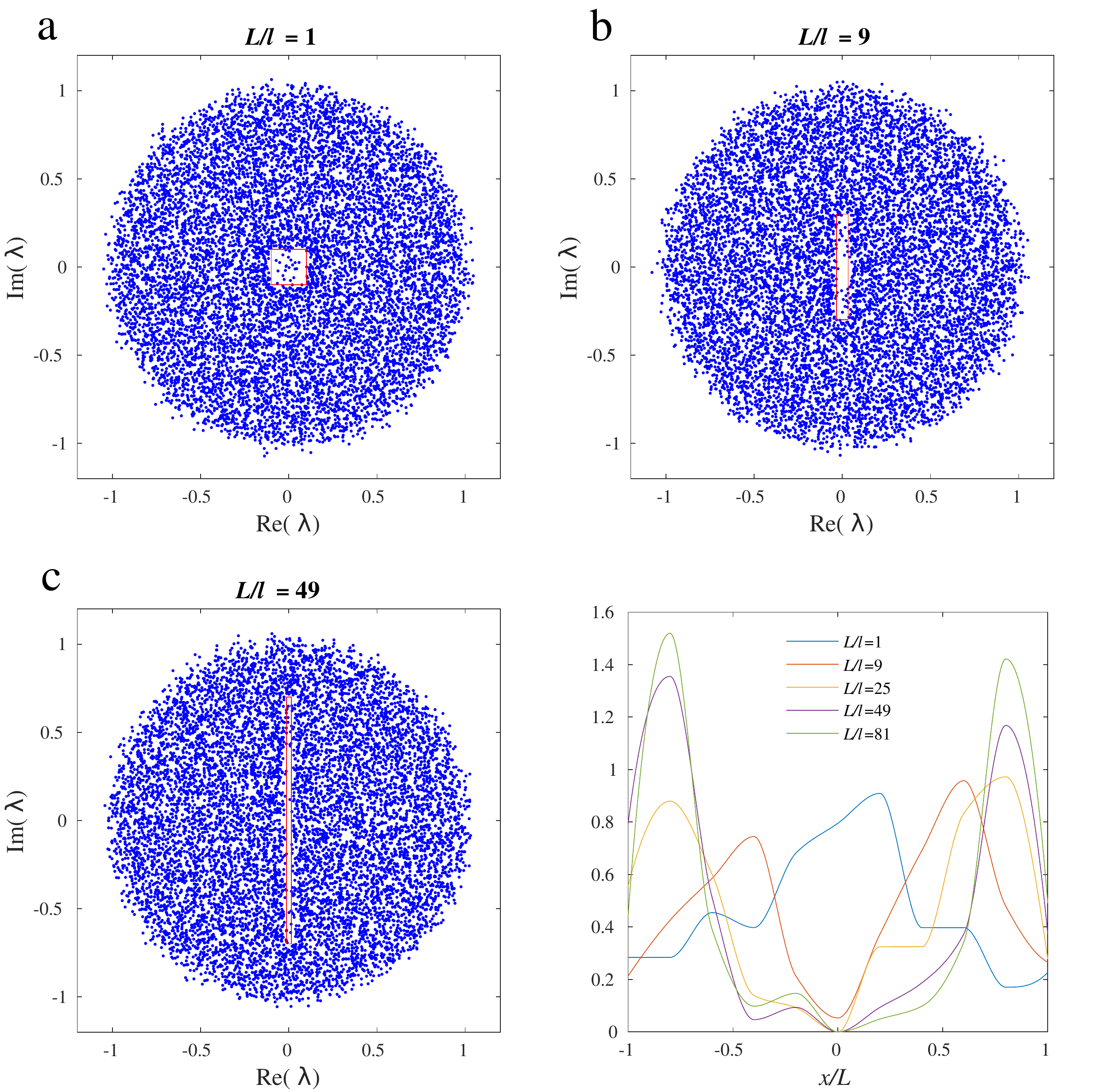}
\caption{\textbf{Conditioning on other subsets of $\C$. (a - c)} Conditioned spectra for 3 different ratios $L/l$. Each figure is the superposition of $20$ matrices with $n = 500$. The number of eigenvalues inside the rectangles is $2$, while the expected number in an equivalent area in an unconstrained matrix is $\approx 6.3$ \textbf{(d)} Distribution of eigenvalues along the long direction of the rectangle. }\label{fig:general_shapes}
\end{figure}

We provide here numerical simulations in order to illustrate the fact that the transition we exhibited on the distribution of real eigenvalues $\mr$ is also present when we condition matrices to have a certain number of eigenvalues $h_n$ inside an arbitrary elongated region of the complex plane. Figure~\ref{fig:general_shapes} shows the case of the complex Ginibre matrices conditioned on having a small number of eigenvalues inside rectangles with different length of their long and short sides ($L$ and $l$ respectively), such that the surface $Ll$ is fixed (thus fixed expected number of eigenvalues) but the ratio $L/l$ changes. We observe that when $L=l$ there is no bimodal distribution, however, as one of the sides become larger than the other the distribution of eigenvalues along the long axis is bimodal. Furthermore, rescaling with respect to the length of the long axis, the peaks of the bimodal distribution converge to the same position.

The presence of the transition for macroscopic regions (i.e. $L,l=\O{1}$) is a finite size effect, but as long as $l=\o{1}$ the log-gas arguments provided before hold and therefore the phenomenon still happens in the large $n$ limit.

\section{Numerical estimations of ESD and $\mr$}
\begin{figure*}
\includegraphics[width=\textwidth]{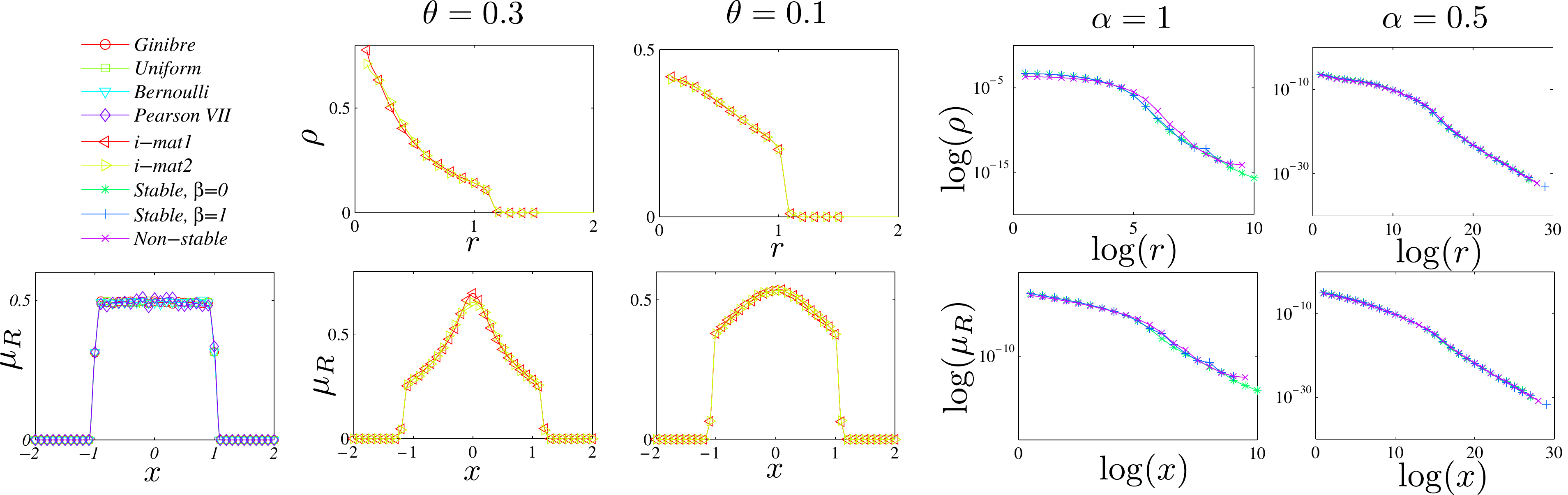}
\caption{\textbf{Numerical estimations of ESD and $\mr$. Top:} Radial distribution of eigenvalues $\rho(r)=\int_0^{2\pi}\mu(re^{i\theta})\frac{d\theta}{r}$ for \im and \h matrices with $n=1000$. \textbf{Bottom:} Distribution of real eigenvalues over the real line $\mu_R$ for \iid, \im and \h matrices with $n=1000$. All \iid matrices converge to the uniform distribution. In the large $n$ limit the empirical spectral distribution of \im and \h matrices depends only on the parameters $\alpha$ and $\theta$. }\label{fig:distributions}
\end{figure*}

For the sake of completeness, in Figure~\ref{fig:distributions} we provide simulations of radial distribution of the ESD (which is asymptotically rotationally invariant), together with the distribution of real eigenvalues, for different instances of matrices in all three families analysed in the main text. We observe that the ESD in the case of \im matrices only depends, in our simulations, on the corresponding $\theta$. The eigenvalues concentrate more densely around the origin of the complex plane than in the circular law, and this is also visible in the unimodal shape of the distribution of real eigenvalues. In the \h case, the radial shape of the distribution appears to only depend upon the tail of the elements distribution $\alpha$ as predicted in \cite{bordenave11}. 

\end{document}